# Atomic-Scale Ordering of Sulfur Vacancies Enhances Charge Transport in Monolayer $MoS_2$

Alessandro Pecchia,*[1] Andrea Lorenzoni,[2] Alexander Croy,[3] Francesco Mercuri,[2] and Massimiliano Cavallini*[2]

1- Istituto per lo Studio dei Materiali Nanostrutturati (ISMN) - Consiglio Nazionale delle Ricerche (CNR) Prov.le 35d n. 9, 00014 Roma, Italy.

2- Istituto per lo Studio dei Materiali Nanostrutturati (ISMN) - Consiglio Nazionale delle Ricerche (CNR) Via P. Gobetti 101, 40129 Bologna, Italy.

3- University of Jena, Room 105, Lessingstraße 4, 07743 Jena, Germany

**Abstract**

Defect engineering in two-dimensional semiconductors has primarily focused on controlling the nature and concentration of atomic defects. Here, we show that the spatial arrangement of defects can be equally decisive in determining electronic transport. Using sulfur vacancies in monolayer $MoS_2$ as a model system, we investigate the impact of vacancy ordering through density functional theory, density functional tight-binding calculations, and quantum transport simulations. We demonstrate that a periodic vacancy arrangement at a concentration of 11.1% transforms isolated defect states into a narrow dispersive in-gap miniband, whereas randomly distributed vacancies generate only localized electronic states. This electronic transition fundamentally alters charge transport, enabling band-like propagation through the defect network rather than transport limited by disconnected localized states. A systematic analysis of the complete symmetry-reduced ensemble of 94 non-adjacent four-vacancy configurations shows that the ordered pattern lies within a broad low-energy manifold and is not energetically anomalous, although it is not the thermodynamic ground state. Device-level simulations of Au/$MoS_2$/Au junctions reveal efficient alignment of the metal Fermi level with vacancy-derived states, promoting charge injection into the defect miniband. As a result, ordered vacancy arrays exhibit electrical currents up to five orders of magnitude higher than statistically equivalent random distributions and can approach, or locally exceed, the transport performance of pristine $MoS_2$. These findings establish atomic-scale defect ordering as a powerful design principle for two-dimensional materials, demonstrating that the organization of defects, beyond their concentration alone, provides a route to simultaneously preserve functionality and high electrical conductivity in highly defective semiconductors.

## Introduction

Atomic defects such as heteroatoms and vacancies play a central role in determining the physical properties of two-dimensional transition metal dichalcogenides (TMDs) [1,2]. They provide opportunities to modify behavior beyond that of pristine materials [1,3,4]. Incorporating such atomic-scale imperfections enables diverse functionalities applicable to technologies ranging from optoelectronics and quantum devices to thermoelectricity and catalysis. These functionalities basically stem from defects acting as dopants, introducing localized electronic states within the bandgap [5–8]. Unlike conventional electronics, which require dopant levels in the parts-per-million, many of these applications demand defect concentrations many orders of magnitude higher, sometimes exceeding 10%, as in catalysis. At such concentrations electrical conductivity is unavoidably reduced. [9,10]

Defect engineering is the means of optimizing defective materials for target applications [11,12], however, research and technology in this field are focused on controlling defect nature and their statistical distribution. This control is not sufficient to exploit the full potential offered by defects.

Here we demonstrate that the arrangement of defects is as important as their concentration, demonstrating that also for high defect concentration an ordered distribution of atomic vacancies let the preservation of electrical properties of pristine materials. As model material, we choose molybdenum disulfide ($MoS_2$) as it is one of the most extensively studied layered materials and serves as a benchmark platform for investigating defect physics and chemistry.  We focus on atomic sulfur vacancies ($V_s$), which are the most abundant and thermodynamically stable atomic defects in $MoS_2$ and are intrinsically present in virtually all manufactured material. Their electronic character, however, is not fixed: environmental exposure and fabrication routes can modify their doping state (e.g., *p* vs. *n* behavior), thereby affecting the optical and charge transport properties [13,14].

Despite the ability of $V_s$ to introduce or enhance functionalities, vacancies are often detrimental for charge transport, as they act as scattering centers [15,16]. Experimental and theoretical studies consistently report a degradation of conductivity with increasing defect concentration, while also pointing to defect engineering, namely, control over defect density and spatial distribution, as the route to mitigate these effects [9,10]. On the other hand, $V_s$ can act as hopping centres that activate new percolative channels for charge transport, which, once optimised, can reach excellent performance [17–19]. Although this regime leads to measurable improvements in conductivity, increasing the vacancy density is constrained by structural stability. As a result, the optimization of random defect distributions provides only a partial solution: the number and continuity of conductive pathways remain limited, and transport properties are still sensitive to local $V_s$ density variations.

We propose atomic-scale ordering of atomic vacancies as the solution to the problem of electrical conductivity and nanoscale uniformity. By imposing a periodic arrangement of $V_s$ with optimized spacing, it is possible to create well-defined percolative channels for charge transport (Figure 1b), overcoming the limitations associated with random distributions. This concept provides a pathway to enhance conductivity while maintaining structural stability and nanoscale uniformity of defective $MoS_2$.

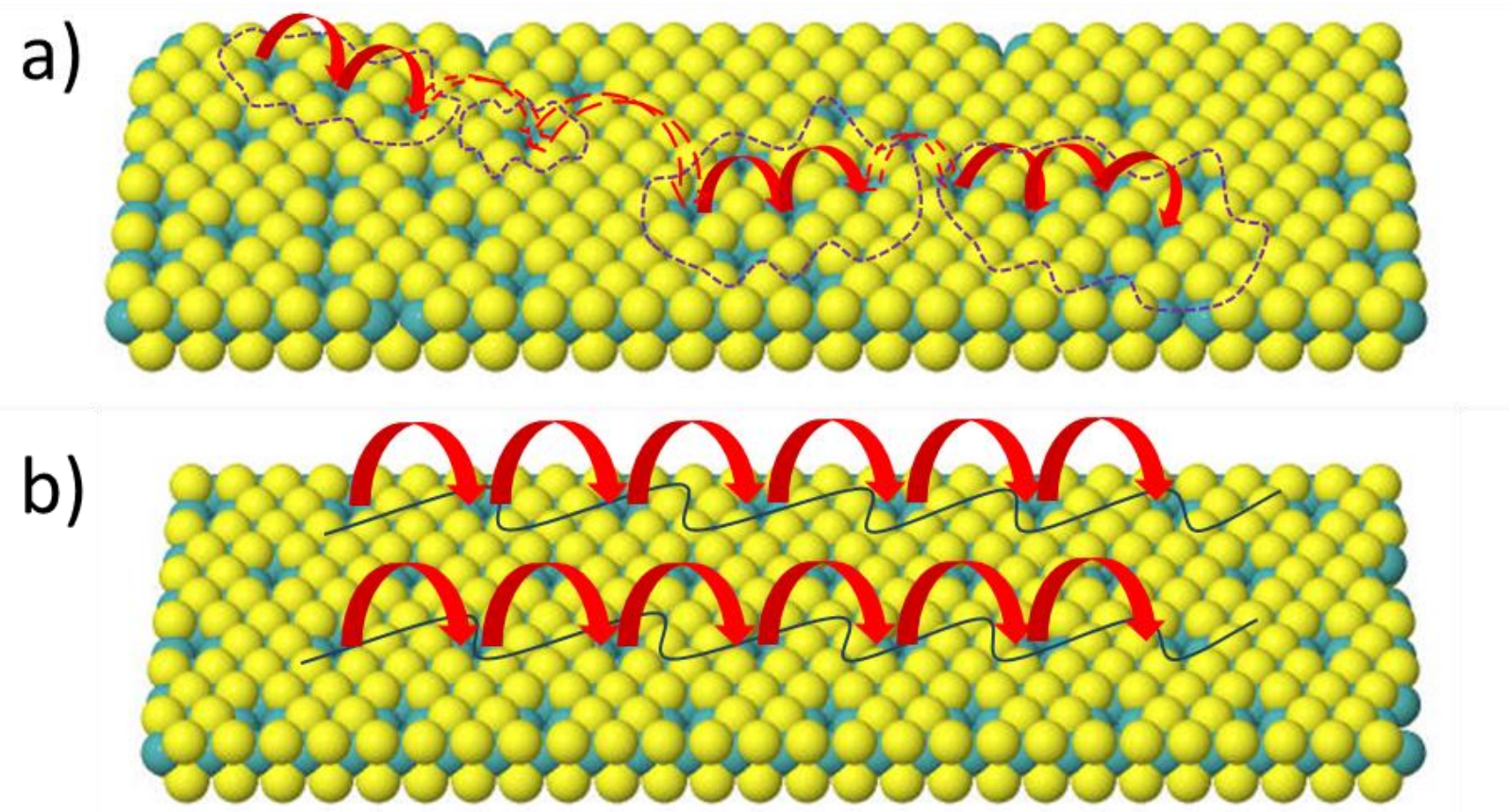

**Figure 1.** *Concept of defect distribution effects in hopping and band charge transport. a) Random percolative transport with hopping between localized states. b) Pathways generated by optimized atomic-scale defect ordering with coherent band-like propagation.*

Clearly, such an ordered configuration cannot form spontaneously but can be obtained using advanced surface nanopatterning techniques [20].

However, a periodic pattern that produces favorable electronic states may nevertheless be energetically disfavored relative to disordered configurations at the same defect concentration. Establishing its energetic position is therefore essential to determine whether it represents a physically plausible ordered model rather than an idealized configuration selected solely for its transport properties. To investigate these aspects, we combine density functional tight-binding (DFTB), targeted density functional theory (DFT) validation, and quantum transport calculations based on Green's function techniques. We first evaluate the energetic position of the ordered pattern within the complete symmetry-reduced ensemble of non-adjacent vacancy configurations at the same defect concentration. We then examine how vacancy ordering modifies the electronic structure and charge transport relative to random configurations.

This work highlights defect ordering as a simple yet powerful design principle for controlling charge transport in two-dimensional semiconductors, underscoring the importance of defect distribution in determining the material properties. Rather than increasing computational complexity, our approach emphasizes a physically intuitive route to overcome one of the main limitations of defective $MoS_2$, opening new directions for nanoscale device engineering.

## Results and Discussion

In defective $MoS_2$, many functional properties arise from the formation of midgap states [21–23], which considerably modify the electronic structure of the material. The localized wavefunctions that extend out to about 8 Å, uniformly in the lattice directions. This implies that when $V_s$-$V_s$ distance exceeds 16 Å [18], the corresponding wavefunctions do not overlap; conversely, below this distance, defect states hybridize, allowing electrons to flow through [18,22].

For the ordered distribution, we selected a 3x3 hexagonal arrangement (Figure 2c). In this structure, the $V_s$-$V_s$ distance is 0.95 nm, with a defect percentage of 11.1 %. We selected this mean distance because it is experimentally accessible and has been proven to be crucial in devices based on defective $MoS_2$, which experimentally exhibit the best-reported properties in charge transport thanks to the spontaneous formation of percolative patterns [17–19]. The ordered structure was compared with random configurations, generated in 6x6 hexagonal supercells containing four vacancies on the same sulfur surface, thereby preserving the same vacancy concentration, excluding adjacent vacancies Representative random configurations were used for the electronic-structure comparison in Figure 2, whereas the complete symmetry-reduced configurational ensemble is analyzed below.

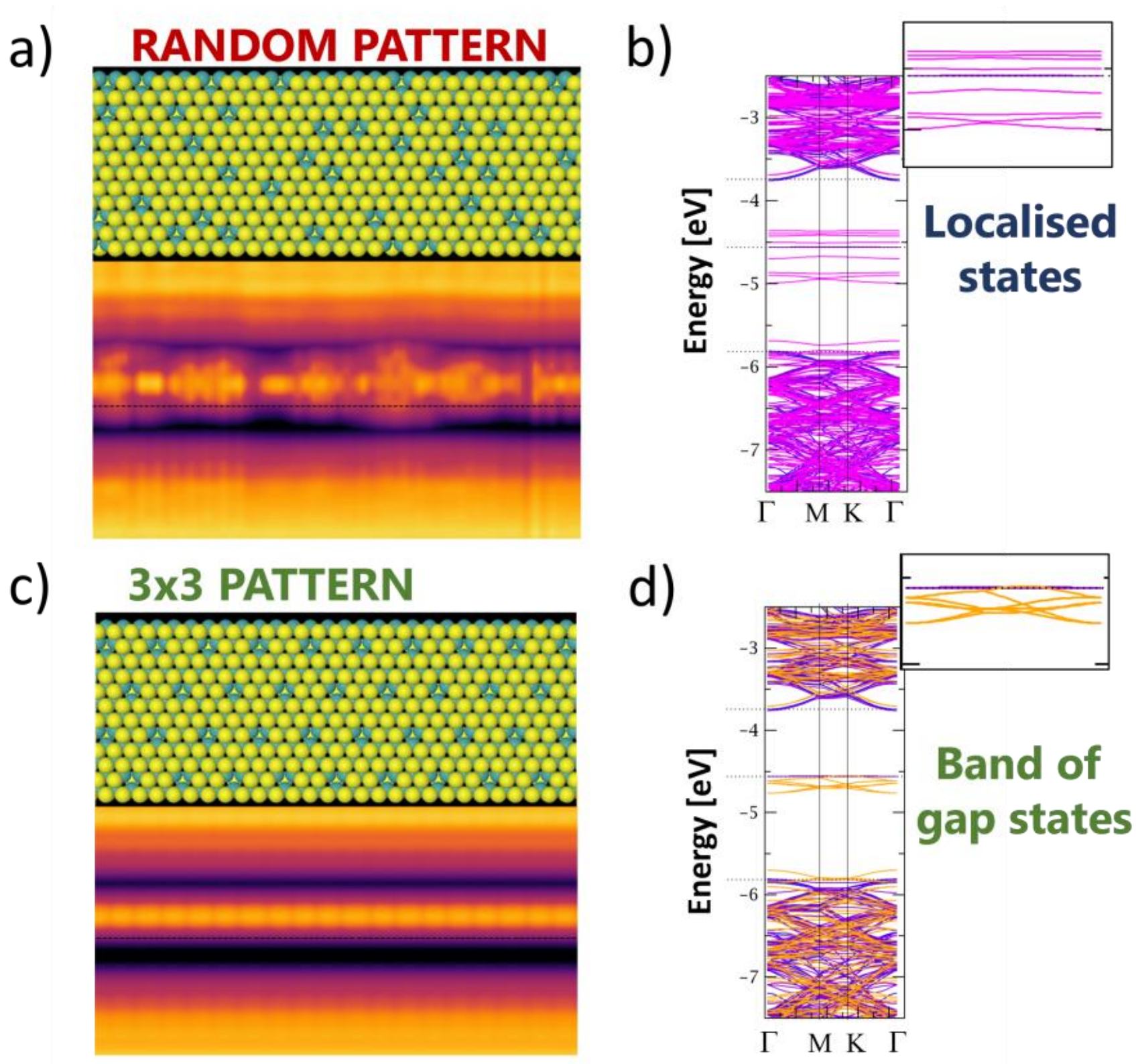

**Figure 2.** *Distribution of vacancy defects within a 8x3 nm² area (11.1% of Vs). a) Example of a Random structure and Projects DOS around the bandgap. b) Bandstructure for a 6x6 unit cell with localized defect states. c) 3x3 ordered distribution and PDOS. d) Bandstructure for a 6x6 unit cell with 3x3 ordered defect pattern. In both b) and d) the bands in the background (purple) correspond to the same cell with 1 defect.*

Random structures were generated using a Python script that removes isolated sulfur atoms from the top surface of the ideal lattice while enforcing the non-adjacent constraint. All 6x6 supercells contain four $V_s$ defects and therefore have the same vacancy concentration as the ordered arrangement. The resulting dataset captures the statistical variability inherent in disordered defect landscapes and provides a robust reference for evaluating the energetic and electronic implications of vacancy ordering. More information and the python script used can be found in the Methods section and in the Supporting information.

Figures 2 show the energy level distribution and the local density of states (PDOS) for both scenarios. In the case of random defects, only localized states are observed within the band gap, positioned approximately 0.9 eV below the conduction band (Fig. 2a-b). In contrast, the ordered defect distribution produces a narrow band of gap states approximately 130 meV wide in the band gap region (Fig.2c-d). The PDOS shown in Figure 2a and 2c is obtained by slicing the system in the long axis direction. Notably, whenever local order or low concentration is recovered, the DOS narrows down and closely resembles the ordered distribution. For comparison, the 4x4 hexagonal pattern leads to a much narrower bandwidth of just 15 meV, which is considered too small to sustain band-like transport and was not further investigated.

The formation of a dispersive in-gap band in the designed ordered vacancy pattern provides the electronic basis for the enhanced transport discussed below. Since this pattern is intentionally designed to impose a periodic arrangement of sulfur vacancies, rather than assumed to be the thermodynamic ground state, we evaluated its energetic position within the broader vacancy configurational landscape. The purpose of this analysis is to verify that the designed ordered pattern is not an energetically anomalous arrangement, but a realistic member of the accessible four-vacancy configurations at the same defect concentration. To this end, the designed ordered vacancy pattern was compared with the complete symmetry-reduced set of four-vacancy configurations at the same vacancy concentration. The final dataset contains 94 inequivalent configurations, including the designed ordered pattern itself. All configurations were relaxed using the same DFTB setup, and their final total energies were compared using the designed ordered pattern as the reference, as described in the Methods section. Figure 3 reports the relative energy, $\Delta E = E_i - E_{ordered}$, as a function of the mean vacancy-vacancy distance. The ordered structure is set to $\Delta E$=0 and has the greatest mean vacancy-vacancy distance (9.27 Å), with all pairwise vacancy distances equivalent by periodicity, identifying it as the most uniform four-vacancy distribution within the 6x6 cell. The other configurations are symmetry-inequivalent non-adjacent patterns and generally contain at least one shorter vacancy-vacancy separation.

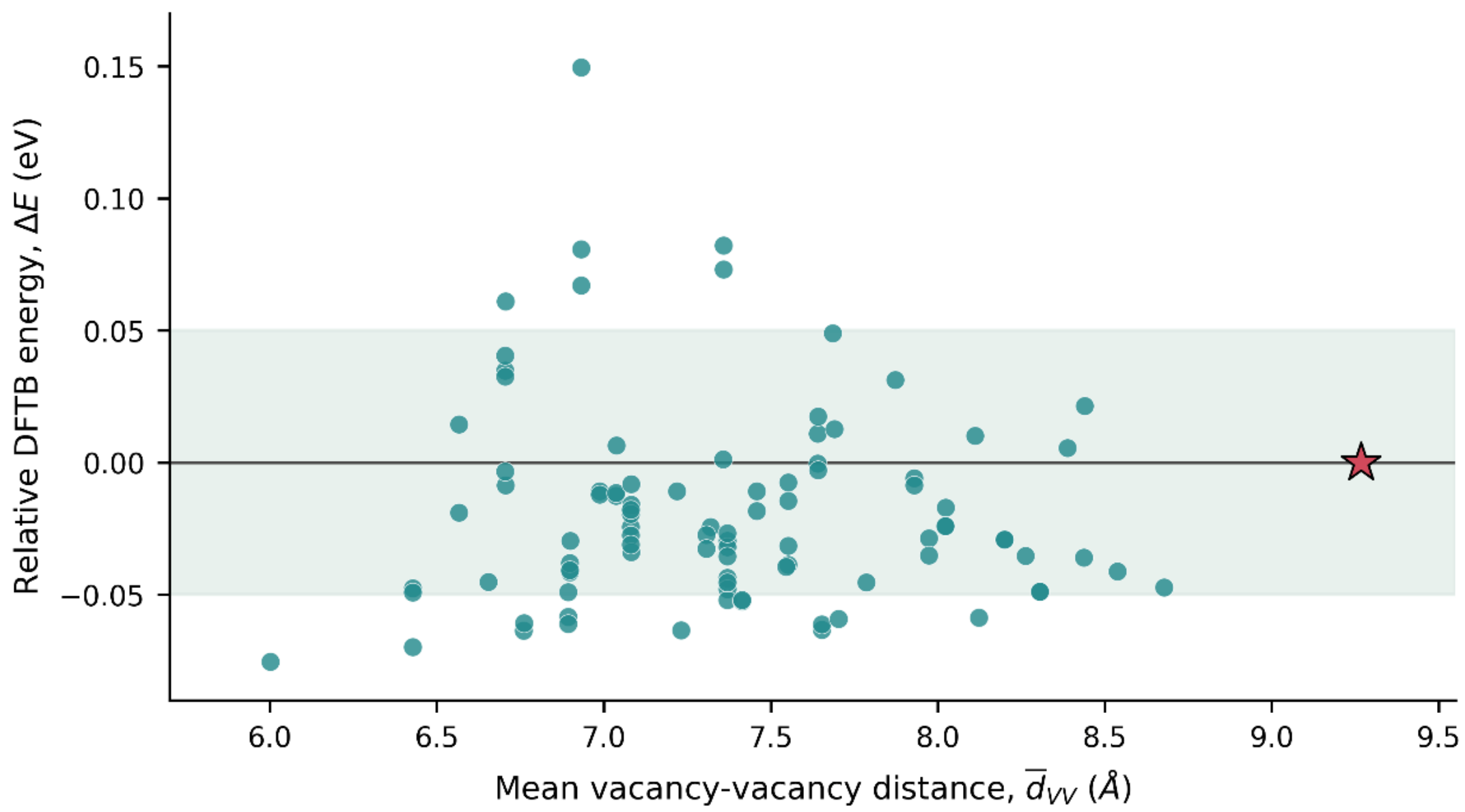

**Figure 3** *Energetic distribution of non-adjacent four-vacancy configurations. DFTB relative energies of the 93 alternative configurations with respect to the ordered 3 × 3 reference are plotted against the mean vacancy–vacancy distance. Each green circle represents one symmetry-inequivalent alternative configuration, and the red star denotes the ordered reference (ΔE = 0). The shaded region marks configurations within |ΔE| < 0.05 eV of the reference.*

The alternative vacancy configurations span the same low-energy region as the ordered model. Among the 93 configurations distinct from the ordered reference, 73 (78.5 %) are quasi degenerate within $|\Delta E| < 0.05$ eV, 14 lie more than 0.05 eV below the reference, and 6 lie more than 0.05 eV above it. The lowest-energy configuration lies 0.0754 eV below the ordered structure, corresponding to approximately 0.019 eV per vacancy. The negligible rank correlation in Figure 3 shows that the mean vacancy-vacancy distance alone does not predict stability within this non-adjacent ensemble. The observed energy variations are instead associated with more local geometric features and with the in-plane relaxation. The ordered pattern should therefore not be interpreted as a unique thermodynamic ground state. Rather, it belongs to a broad low-energy manifold and is an energetically reasonable periodic model for isolating the effect of vacancy ordering. Full energies, structures, descriptors, and symmetry checks are reported in the Supporting Information.

A targeted PBE calculation was performed on eight representative configurations spanning the full DFTB energy range. The resulting relative energies show strong agreement with DFTB (Pearson r = 0.986; MAE =

0.027 eV). Fixed-cell DFT relaxations of the lowest-energy DFTB configuration, the ordered reference, and the highest-energy DFTB configuration preserve the same energetic ordering and produce only small changes respect to the DFTB geometries (RMS atomic displacement of 0.025-0.026 Å). This targeted comparison supports the use of DFTB for configurational trends. Computational details and complete validation results are reported in the Supporting Information.

## Transport across Au – MoS2 –Au resistor

In order to give a more realistic insight into the transport properties of these systems, we model devices in which the $MoS_2$ monolayer is contacted by Au electrodes. We considered crystalline Au (111), which is the most common orientation for thin film contact electrodes. We assume three monolayers of Au on top of a $MoS_2$ monolayer, finding a near lattice-matched configuration that result in a small supercell. It is known that this interface can give rise to Moiree patterns [24], especially when a supercell 11x11 of Au is matched to a 10x10 cell of $MoS_2$. However, this gives rise to a supercell which is way too large for the transport calculation. Since the lattice of $MoS_2$ relaxed with dftb+ is 0.319 nm is slightly larger than the experimental value of 0.317 nm, and the electronic properties of Au do not change significantly if compressed from 0.288 Au-Au distance to 0.276 nm (6.3% compressive strain), we created a model of the interface as shown in Figure 3b. The compressive strain in-plane is relaxed along the (111) direction.

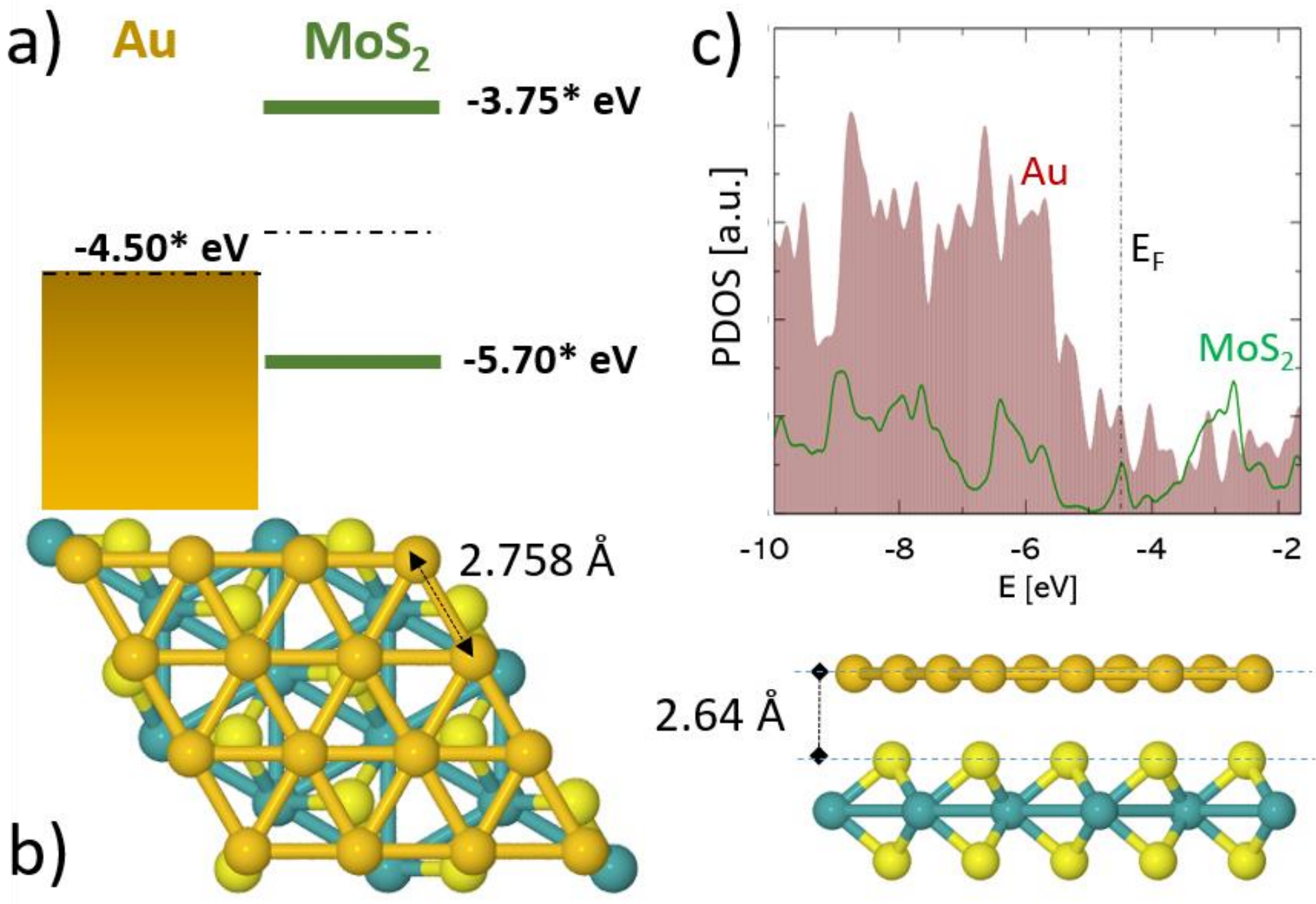

**Figure 3.** *a) Band alignment of separated Au and $MoS_2$. b) Model of the lattice-matched atomic interface between Au 4x4 and $MoS_2$ 5x5 R60°. c) Projected DOS on 3ML Au (brown) and $MoS_2$ (green) with 1 vacancy defect. Au Fermi level aligns to the Vs state.*

The Au-S plane distance is 0.264 nm, which agrees with DFT studies [25]. The non-interacting level alignment between the Au Fermi level and the $MoS_2$ band edges is shown in Figure 3a, reporting the values as computed by the DFTB method. The accepted experimental values for the Ionization Potential (IP) and Electronic Affinity (EA) of $MoS_2$ are about 6.3 eV and 4.3 eV, respectively. These values are 0.6 eV larger than the computed DFTB band edges, which are found at –3.75 eV and –5.70 eV for the conduction and valence band, respectively. Here we assume a reference vacuum level at 0. The underestimate in the IP is due to the approximate description of the electronic density in DFTB, which is described via atomic Mulliken charges, but it also affects DFT calculations with common PBE functionals. In order to obtain a consistent alignment between the Au Fermi level and the MoS2 bands, we have imposed an electrostatic potential to the Au atoms, effectively tuning an effective Au workfunction (WF) to 4.5 eV. In relative terms, taking into account a common shift of –0.6 eV, this would correspond to a real gold WF of 5.10 eV which agrees with the literature (See the SI for a technical discussion). The $MoS_2$ layer contains the 3x3 pattern of vacancy defects discussed above, and the Au/$MoS_2$ interface is left to interact within a self-consistent DFT(B) calculation, resulting in a Fermi level alignment to the Vs states. The final alignment is shown in Figure 3c, where it is possible to see the DOS projected on Au and $MoS_2$ and the resonating alignment of the Fermi level with the defect state. This alignment mechanism is rather robust to small variations of the metal Fermi level as confirmed by changing the Au atom potential shift. Indeed, it is known that Fermi level pinning at metal/$MoS_2$ interfaces is primarily due to defects [26,27]. Despite this, the Au/$MoS_2$ interface is known to have good properties for electronic devices [28,29] and charge injection. Au atoms are known to have a strong interaction with the sulfur vacancies [30], likely indenting in the S voids [31] and Vs are able to modulate the Schottky barrier [32].

The Fermi-level alignment at the defect states provides a significant benefit for electronic injection in the defect bands, providing an efficient pathway for charge transport. To study this, we created suitable models of Au/$MoS_2$/Au with 2-pads contacts, as shown in the lowest diagrams of Figure 4. The models comprise 2,558 atoms, with a contact-to-contact channel length of 10 nm, 3.5 nm wide. We compare the transport properties of ideal non-defected $MoS_2$, the ordered 3x3 pattern and 20 samples of random distributions of defects in the channel between the Au contacts. Both the ordered and disordered systems contain 64 defects each. This was done in order to compare structures with exactly the same defect density. An ordered 3x3 defect pattern is assumed beneath the Au pads. This was done because in the framework of NEGF calculations, the contacts are assumed semi-infinite periodically repeated atomistic structures, a technical scheme that allows to exactly compute the contact self-energies. As a consequence, random defect patterns are in fact periodically repeated, introducing possible artefacts. Additionally, the comparison between different random samples becomes more difficult because the variations are not limited to the defect patterns but also to the contact leads and to a minor extent also the electrostatics and band bending.

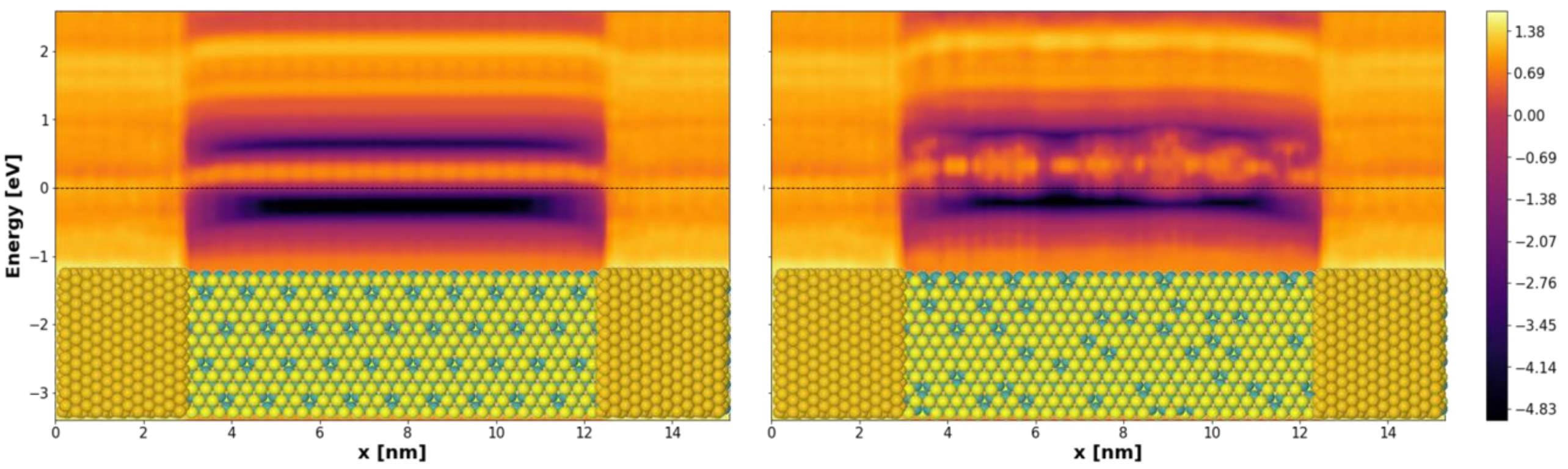

**Figure 4** *Au/MoS2/Au devices and PDOS evidencing the difference between the ordered pattern and a random sample.*

Indeed, figure 4 shows the PDOS for the ordered and one disordered sample. It is possible to notice a band bending resulting from the self-consistent charge-density solution which is induced by the band alignment at the Au/$MoS_2$ interface. The bending can be appreciated comparing with Figure 2, showing the uncontacted layer and flat bands. In moving towards the middle of the channel the band alignment between Au and $MoS_2$ tends to the non-interacting case, depicted in Figure 3a. The resulting long-range electrostatic field is causing convergence issues in the calculations. It is important to point out that the systems shown in Figure 4 do not directly converge. This is not genuinely due to charge oscillations or the known problem of charge sloshing, but rather the charge density quickly diverges unless a good initial guess is provided. Luckily, short channel devices do converge, so the procedure adopted was to first converge a short channel (typically 3 nm) and then progressively increasing the channel length by adding atoms and patching the previously converged charge density for the next run. We could not find any better strategy than this rather lengthy procedure, although assisted by scripts.

As shown in Figure 5, the in-gap transmission through the defect states indicates defect-mediated transport, with percolation through the connected defect states, both in the ordered and the random scenarios. This contribution is not present in pristine $MoS_2$. The ordered pattern mainly shows a significantly higher value of electron transmission, particularly in the energy range [-4.4, -4.1] eV, corresponding to the energy of the defect states or bands (see Figures 2 and 4). All transmissions shown in Figure 5 have been computed in equilibrium condition.

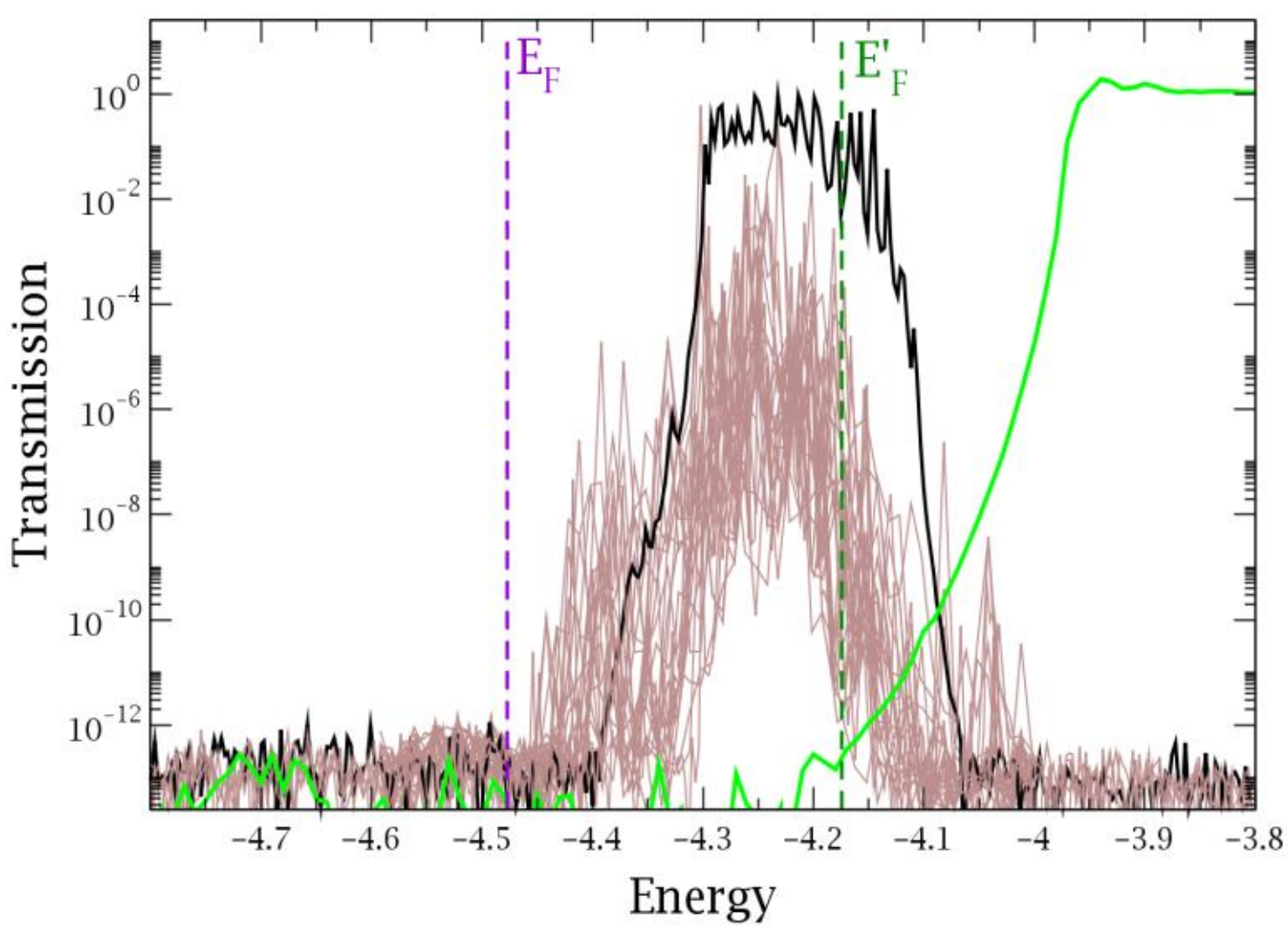

**Figure 5.** *a) Transmission spectra for the ordered structure 20 random structures with 11.1% of sulfur vacancies. The Black curve corresponds to the ordered defect pattern, the thin brown curves correspond to the random distributions, green line to the ideal undefected* $MoS_2$*. The vertical dashed lines represent the Fermi level of the defected (*$E_F$*) and the ideal (*$E'_F$*) systems.*

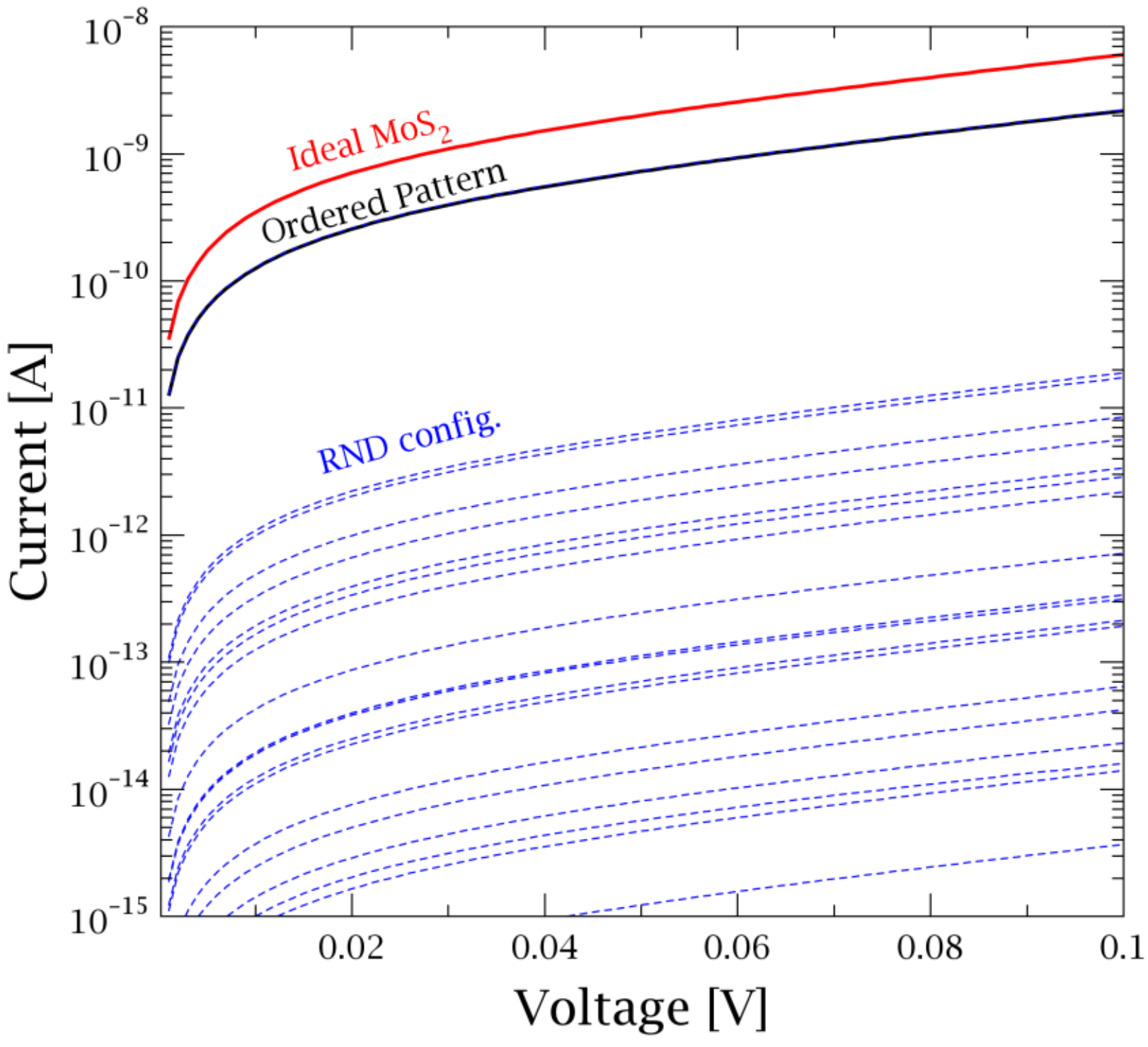

**Figure 6.** *I–V curves for the ideal undefected* $MoS_2$ *(red),* $MoS_2$ *containing ordered defect distribution (black) and Random systems (blue dashed). The plot has been obtained for T=300 K.*

The I-V characteristics shown in Figure 6 have been obtained by imposing a small bias to the system up to 100 meV. The striking result is that the ordered pattern gives the best current by several orders of magnitude compared even to the best disordered sample. The currents obtained for the random configurations span several orders of magnitude, showing that the current is quite sensitive to the defect arrangement. The highest currents are obtained when a contact-to-contact percolation pattern is established. The ordered pattern almost recovers the I-V characteristic of the ideal $MoS_2$. However, the latter is sensitive to the injection barrier for electrons, as can be seen in Figure 5, where the Fermi level is found about 250 meV below the conduction band edge, but this value can change depending on several parameters. The Schottky barrier at metal/$MoS_2$ interfaces is a known problem and usually Au/Ti is employed to reduce the metal workfunction and improve metal/layer injection, but even in this improved experimental condition, a Schottky barrier is known to be formed at the interface, limiting the current in $MoS_2$ devices.

These results demonstrate a significant increase in transport for $MoS_2$ layers with ordered defects patters, reaching up to 5 orders of magnitude improvement compared to random configurations (3 orders of magnitude for the best configuration). This impressive enhancement can be attributed to the delocalization of charge carriers facilitated by mini-band formation in the ordered defect structures, as observed in the band structure analysis. Fermi level tuning that might be obtained for instance using Au/Ti metal contacts could further improve injection in the defect band, also avoiding the band bending observed in our calculations, leading to a significant increase of current. Indeed, considering the transmission close to 1.0 for the ordered band, as shown in figure 5 (peak in the range [–4.30, –4.15] eV), one might expect a current close to 1 μA at 100 meV.

## Conclusions

In this work, we demonstrate that the arrangement of defects is as important as their concentration in determining the performance of two-dimensional materials. In monolayer $MoS_2$, periodic atomic-scale ordering of sulfur vacancies can counteract the transport degradation generally associated with high vacancy concentrations.

By combining density functional tight-binding calculations and quantum transport simulations, we investigated the energetic, electronic, and transport consequences of vacancy ordering. Our findings reveal three fundamental points in defect engineering: first, an ordered vacancy arrangement promotes the hybridization of localized defect states into a dispersive in-gap miniband, providing a connected pathway for band-like charge transport; second, the ordered configuration lies within a broad low-energy manifold of non-adjacent vacancy arrangements and is therefore not energetically anomalous; third, in the investigated Au/$MoS_2$/Au devices, the ordered vacancy array produces electrical currents three to five orders of magnitude higher than the representative random configurations. This enhancement compensates for the reduction in transport associated with randomly distributed vacancies and may reduce the operating bias required to achieve a given current.

Although our study focuses on $MoS_2$ as a representative system, we emphasize that the underlying principle may extend to a wide range of systems beyond $MoS_2$. This highlights the potential of atomic-scale defect ordering and identifies defect arrangement as an emerging design objective in defect engineering.

The experimental realization of artificial atomic-scale defect ordering in TMDs remains a challenge. However, recent advancements in defect engineering, such as the precise fabrication of single atomic vacancies, advanced defect nanopatterning techniques, and the phenomenon of spontaneous ordering, provide promising routes towards this objective in the near future.

## Methods

### DFTB framework and quantum-transport calculations

The calculations presented in this work have been performed with the dftb+ code [33], which implements the Density Functional Tight Binding method [34]. The reason for this choice is mostly connected to the transport calculations based on Green's Functions for which we have generated rather large samples of 2558 atoms, impractical for ab-initio approaches.

The DFTB method is well-established in the literature and is known to provide accurate results of near DFT quality in terms of geometries, heat of formation as well as vibrational properties, provided a good material parameterization is available.

The DFTB method is based on Slater-type minimal basis set of orbitals computed at DFT level for atomic centers with an empirical compressive function of the form $(r/r_0)^\alpha$. In the tight-binding spirit, only the two-center integrals in the Hamiltonian matrix elements between local basis functions are retained and tabulated. The ion-ion and electronic repulsive energies are not computed directly from first principles but fitted against DFT total energies of reference systems. For $MoS_2$ we have closely followed Heine's periodic table recipes [35]. A comparison of the electronic bandstructure against Quantum Espresso calculations is discussed in the SI. Beside the electronic properties, we also assessed relaxed geometries, phonon bandstructure and lattice relaxation around the defect and found a good agreement with DFT calculations.

The original DFTB method has been extended to include self-consistent charge density, spin density, LDA+U, excited state dynamics, non-adiabatic dynamics and, more recently, range-separated functionals [36]. The extension to Non-equilibrium Green's functions (NEGF) was also accomplished long time ago [37,38]. All these developments have made DFTB one of the most successful semi-empirical approach in condensed matter theory.

The defected structures have been created based on the starting from the unit cell relaxed using Quantum Espresso, norm-conserving pseudopotentials available at https://www.pseudo-dojo.org, PBE functional and an energy cutoff of 70 Ry. Atoms have been removed either in the ordered pattern or with a randomized scheme. We removed S atoms only from one face and intentionally prevented nearest neighbors vacancies because these configurations are considered rare in the literature and can strongly bias the statistics performed on small supercells. Geometry optimization utilized conjugate gradient algorithms until atomic forces fell below the convergence threshold. Convergence criteria were set to achieve self-consistent solutions for electronic structure calculations, with thresholds for energy and force convergence set at $10^{-5}$ Hartree and $10^{-3}$ Hartree/Bohr, respectively.

## Configurational stability analysis of vacancy arrangements

For the systematic configurational-stability analysis, starting from the Quantum ESPRESSO-relaxed unit cell described above, a 6×6 $MoS_2$ monolayer supercell was constructed and four sulfur atoms were removed from the exposed sulfur layer. The four vacancies correspond to 11.1% of the 36 sites in that surface sublattice. The ordered 3×3 vacancy pattern investigated in the electronic-structure and transport calculations was included as the reference configuration.

All 58,905 four-site subsets of the 36 exposed-surface sulfur sublattice were generated combinatorially. Configurations were retained only when every vacancy-vacancy distance exceeded 3.5 Å under periodic boundary conditions, thereby excluding adjacent-vacancy motifs. This filtering yielded 16,659 valid arrangements.

Symmetry-equivalent configurations were identified using all 36 lattice translations and the six C3v operations that preserve the exposed sulfur sublattice, corresponding to the D3 action on the two-dimensional vacancy-site lattice. Operations exchanging the upper and lower sulfur planes were not considered. The symmetry reduction produced 94 inequivalent configurations. An orbit-size validation confirmed that the degeneracies of the representatives account exactly for all 16,659 retained arrangements. The resulting 94 structures were then optimized with SCC-DFTB using the same Mo-S Slater-Koster parameterization described above and identical numerical settings. For this dedicated configurational dataset, a stricter geometry-convergence criterion was adopted than that used for the initial transport structures. Conjugate-gradient optimization was performed for up to 4,000 steps, with the two in-plane lattice-vector lengths allowed to relax independently. The 20 Å out-of-plane cell length and all cell angles were kept fixed. The maximum force-component threshold was $1.0 \times 10^{-5}$ Hartree/Bohr, and the SCC tolerance was $1.0 \times 10^{-6}$, with a maximum of 100 SCC iterations. Brillouin-zone sampling employed a shifted 4×4×1 k-point grid, and electronic occupations were described using Fermi filling at 100 K. The maximum angular momenta were set to *d* for Mo and *p* for S.

Configurational energies were expressed relative to the ordered pattern according to

$$\Delta E = E_i - E_{ordered}$$

where $E_i$ is the final optimized DFTB total energy of configuration $i$, and $E_{ordered}$ is that of the ordered 3×3 reference. Negative $\Delta E$ values therefore identify configurations below the ordered reference, whereas positive values identify configurations above it. All structures have the same composition, containing 36 Mo and 68 S atoms, so the comparison is performed at fixed vacancy concentration. The interval $|\Delta E| < 0.05$ eV was used as an operational criterion to group configurations close in energy to the ordered model and should not be interpreted as an estimate of the numerical uncertainty.

Geometric descriptors were calculated from the optimized structures. The six vacancy-pair distances were evaluated using the minimum-image convention and used to determine the minimum, mean, maximum, and standard deviation of the vacancy-vacancy separation. Additional descriptors included the in-plane cell-area strain, individual lattice-vector strains, sulfur-layer thickness, vacancy radius of gyration, vacancy-distribution anisotropy, and atomic displacements relative to the corresponding unrelaxed defective structures. Pearson and Spearman coefficients were used to evaluate linear and monotonic relationships, respectively, between these descriptors and the relative configurational energy. Complete descriptor definitions and values for all 94 structures are reported in the Supporting Information.

### Targeted DFT validation of configurational energetics

Single-point DFT calculations were performed on DFTB-relaxed geometries. The DFT calculations were performed with Quantum ESPRESSO using spin-polarized PBE and PAW pseudopotentials for Mo and S. Wavefunction and charge-density cutoffs of 50 and 400 Ry were used, respectively. Reciprocal-space sampling employed a shifted 2×2×1 Monkhorst-Pack grid, and electronic occupations were treated using Marzari-Vanderbilt smearing with a width of 0.005 Ry. The electronic convergence threshold was $1.0 \times 10^{-8}$ Ry, the mixing factor was 0.2, and up to 250 electronic iterations were allowed. Spin polarization was enabled with an initial Mo magnetization of 0.1. Symmetry and inversion operations were disabled to preserve each vacancy arrangement.

The three ionic optimizations were performed using the BFGS algorithm while keeping the DFTB-optimized simulation cells fixed. DFT and DFTB relative energies were independently referenced to the ordered configuration. Complete input files, convergence information, relative-energy tables, and structural-displacement analyses are provided in the Supporting Information.

## Conflict of Interest

The authors declare no conflict of interest.

## Author Contributions

M.C. conceived the work and wrote the manuscript. A.P. performed the calculation of energy band structure and charge transport. A.L. and F.M., performed the calculation of system stability, A.C. performed the DFTB parameterizations. All authors discussed the results, commented and revised on the manuscript and all authors approved the final version before submission to the journal.

**Data Availability Statement**

The data supporting this study's findings are available from the corresponding author upon reasonable request.

## Acknowledgements

We thank the European Union funded this research – Next Generation EU from the Italian Ministry of University and Research. Project PRIN 2022SRHPF2 ”Molecular assisted atom vacancies arrangement to modulate magnetism in 2D transition metal dichalcogenides” (MAVAM) and Italian Ministry of Environment and Energy Security POR H2 AdP MMES/ENEA with involvement of CNR and RSE, PNRR – Mission 2, Component 2, Investment 3.5 ”Ricerca e sviluppo sull'idrogeno” under the ENEA – UNIMIB and ENEA-CNR agreements (Procedure 1.1.3 PNRR POR $H_2$). A.P thanks the EuroHPC project “EoCoE-III” GA-101144014.